# Fourier analysis of RGB fringe-projection profilometry and robust phase-demodulation methods against crosstalk distortion

**Moises Padilla**[*], **Manuel Servin** and **Guillermo Garnica**

*Centro de Investigaciones en Optica A.C., Loma del Bosque 115, Leon, Guanajuato, Mexico*
[*]*moises@cio.mx*

**Abstract:** In this paper we apply the frequency transfer function (FTF) formalism to analyze the red, green and blue (RGB) phase-shifting fringe-projection profilometry technique. The phase-shifted fringe patterns in RGB fringe projection are typically corrupted by crosstalk because the sensitivity curves of most projection-recording systems overlap. Crosstalk distortion needs to be compensated in order to obtain high quality measurements. We study phase-demodulation methods for null/mild, moderate, and severe levels of RGB crosstalk. For null/mild crosstalk, we can estimate the searched phase-map using Bruning's 3-step phase-shifting algorithm (PSA). For moderate crosstalk, the RGB recorded data is usually preprocessed before feeding it into Bruning's PSA; alternatively, in this paper we propose a computationally more efficient approach, which combines crosstalk compensation and phase-demodulation into a single process. For severe RGB crosstalk, we expect non-sinusoidal fringes' profiles (distorting harmonics) and significant uncertainties on the crosstalk calibration (which produces pseudo-detuning error). Analyzing these distorting phenomena, we conclude that squeezing interferometry is the most robust demodulation method for RGB fringe-projection techniques. We support our conclusions with numerical simulations and experimental results.

## 1. Introduction

Multichannel red, green and blue (RGB) color encoding is a very useful approach in fringe projection profilometry, which allows simultaneous acquisition of three linearly independent phase-shifted fringe patterns for single-shot measurements [1-6]. Pioneer applications of RGB fringe-projection are due to Wust and Capson [1], who in turn used the phase-demodulation methods by Bruning et al. [7] and Takeda et al. [8]. In RGB fringe-projection profilometry, a digital projector casts simultaneously up-to three sinusoidal phase-shifted fringe patterns over the tridimensional (3D) object under test, and these fringe patterns became phase-modulated proportionally to the height profile of the object [1-6]. Under ideal conditions, we can estimate the searched modulating phase from the RGB encoded data using Bruning's 3-step phase-shifting algorithm (PSA) [1,7]. However, most color cameras and digital projectors have overlapping spectral sensitivities, which produce RGB-crosstalk distortion [9]. Phase-shifted fringe patterns distorted by RGB crosstalk have non-sinusoidal profiles and their phase-steps differ from the nominal value, which is particularly challenging for few-step algorithms [10-12]. In order to improve the quality of the measured phase proportional to the object's profile under study, crosstalk compensation is required before feeding the data into the 3-step PSA [1-6]. Other common practices in fringe projection include gamma correction and projection defocusing in order to obtain more sinusoidal profiles [1-6,13-15].

In this paper, we employ the frequency transfer function (FTF) formalism to analyze the robustness of Bruning's 3-step PSA when applied to fringe patterns distorted by RGB crosstalk. From our analysis, we deduce an equivalent but computationally more efficient method to perform the quadrature filtering in fringe patterns with mild to moderate RGB crosstalk. We also show that Bruning's 3-step PSA is unreliable for severe RGB crosstalk, since it introduces pseudo-detuning distortion and preserves too many distorting harmonics. When working under these conditions we strongly recommend working with the squeezing interferometry technique (which we also analyze briefly in this paper) for robust quadrature filtering. Finally, we support our conclusions with numerical simulations and by digitizing a static 3D surface (for repetitively, in order to compare all methods discussed in this paper).

## 2. Theoretical background

*2.1 The FTF formalism and Bruning's 3-step PSAs.*

For ease of exposition and the reader's convenience, let us begin with a brief review of the FTF formalism applied to phase-shifting algorithms [12]. Our standard mathematical model for a temporal-carrier fringe pattern is given by:

$$I(x,y,t) = a(x,y) + b(x,y)\cos[\varphi(x,y) + \omega_0 t], \qquad (1)$$

where $a(x,y)$ represents the background signals, $b(x,y)$ is the fringes' contrast function, and $\varphi(x,y)$ is the searched modulating phase. Finally, the temporal carrier frequency $\omega_0 \in (0, \pi)$ is assumed to be known. Taking the temporal Fourier transform of Eq. (1), we have:

$$\mathcal{F}\{I(x,y,t)\} = I(x,y,\omega) = a(x,y)\delta(\omega) + \frac{1}{2}b(x,y)\left[e^{i\varphi(x,y)}\delta(\omega-\omega_0) + e^{-i\varphi(x,y)}\delta(\omega+\omega_0)\right]. \quad (2)$$

We need to isolate one of the spectrally displaced analytic signals (without loss of generality we choose the one at $\omega = \omega_0$) in order to demodulate the searched phase. This is done using an $N$-step frequency transfer function $H(\omega)$ that fulfills the quadrature conditions:

$$H(\omega) = \sum_{n=0}^{N-1} c_n \exp(-in\omega); \quad (c_n \in \mathbb{C}),$$
$$H(-\omega_0) = H(0) = 0, \quad H(\omega_0) \neq 0. \quad (3)$$

From Eq. (2) and (3), taking $I(x,y,\omega)H(\omega)$ results:

$$I(x,y,\omega)H(\omega) = \frac{1}{2}b(x,y)\exp[i\varphi(x,y)]H(\omega_0)\delta(\omega-\omega_0). \quad (4)$$

Back on the temporal domain we evaluate at $t = N-1$ in order to use all available data [12], and using the compact notation $I_n(x,y) = I(x,y,t)\delta(t-n)$, we have:

$$A_0(x,y)\exp[i\hat{\varphi}(x,y)] = I(x,y,t)*h(t)\big|_{t=N-1} = \sum_{n=0}^{N-1} c_n I_n(x,y), \quad (5)$$

where $A_0(x,y) = (1/2)H(\omega_0)b(x,y)$ and we use a "hat" to label estimated values. Once we isolate this analytic signal, it is easy to solve for all unknowns in Eq. (1) as:

$$\hat{a}(x,y) = \frac{1}{N}\sum_{n=0}^{N-1} I_n(x,y); \quad \hat{b}(x,y) = \frac{2}{|H(\omega_0)|}\left|\sum_{n=0}^{N-1} c_n I_n(x,y)\right|;$$
$$\hat{\varphi}(x,y) \bmod 2\pi = \tan^{-1}\left\{\frac{\text{Im}[\sum c_n I_n(x,y)]}{\text{Re}[\sum c_n I_n(x,y)]}\right\} = \tan^{-1}\left\{\frac{\sum b_n I_n(x,y)}{\sum a_n I_n(x,y)}\right\}, \quad c_n = a_n + ib_n. \quad (6)$$

Since the estimated phase is wrapped within $(-\pi, \pi]$, we generally apply a phase unwrapper as last step of the demodulation process [12,16].

Particularly, in this paper we will work with the 3-step PSA by Bruning et al. [7,12], for which $\omega_0 = 2\pi/3$ and $c_n = \{1, e^{-i2\pi/3}, e^{-i4\pi/3}\}$, resulting:

$$A_0(x,y)\exp[i\hat{\varphi}(x,y)] = \sum c_n I_n(x,y) = I_0(x,y) + e^{-i2\pi/3}I_1(x,y) + e^{-i4\pi/3}I_2(x,y). \quad (7)$$

Since the coefficients $c_n$ also define the FTF, we have after some algebraic manipulation:

$$H(\omega) = \sum c_n e^{-in\omega} = \left[1 - e^{-i\omega}\right]\left[1 - e^{-i(\omega+2\pi/3)}\right]. \quad (8)$$

As illustrated in Fig. (1), this FTF fulfills the quadrature conditions (see Eq. (3)), thus rejecting the spectral components at $\omega = \{0, -2\pi/3\}$ and amplifying the searched analytic signal at $\omega = 2\pi/3$. This is the ideally expected result.

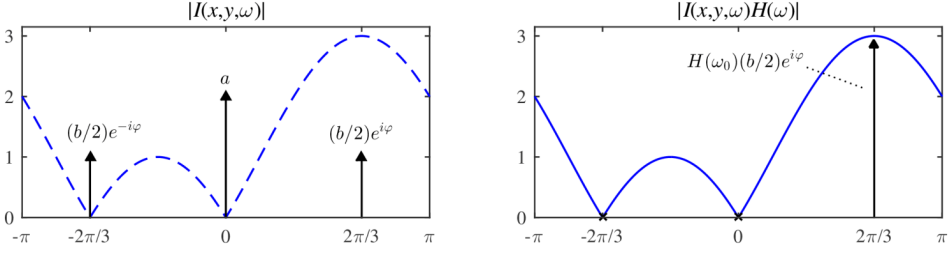

Fig. 1. Fourier spectra of the temporal-carrier fringe pattern before (on the left-side) and after (on the right-side) the quadrature linear filtering given by Bruning's 3-step PSA [7,12].

*2.2 RGB fringe projection profilometry with 3-step phase shifting.*

Three-step phase-shifted fringe projection with a single-frequency spatial carrier is arguably the most commonly used approach in RGB fringe-projection profilometry (for alternative approaches see, for instance, Zhang's review [5] and its references). Our mathematical model for these spatial-carrier open fringes is given by:

$$f_0(x,y) = 255 \times LPF\{[0.5 + 0.5\cos(u_0 x + 0)]^{1/\gamma}\},$$
$$f_1(x,y) = 255 \times LPF\{[0.5 + 0.5\cos(u_0 x + 2\pi/3)]^{1/\gamma}\}, \quad (9)$$
$$f_2(x,y) = 255 \times LPF\{[0.5 + 0.5\cos(u_0 x + 4\pi/3)]^{1/\gamma}\}.$$

Here $LPF\{\cdot\}$ represents an analog low-pass filter (usually defocusing), $u_0$ is the frequency of the spatial carrier (in radians per pixel), and the numerical values for $\gamma$ are within the range $(1.6, 2.4)$. The low-pass filtering and the non-linear gamma encoding allow us to attenuate nonlinear intensity distortions during the fringe-projection process [14,15].

Considering the profilometer setup in Fig. (2), one-at-a-time projection of the open fringes in Eq. (9) produces the following phase-shifted fringe-patterns [1]:

$$I_0(x,y) = a(x,y) + b(x,y)\cos[\varphi(x,y) + 0],$$
$$I_1(x,y) = a(x,y) + b(x,y)\cos[\varphi(x,y) + 2\pi/3]; \quad \varphi(x,y) = u_0 \tan(\theta)[p(x,y) + x], \quad (10)$$
$$I_2(x,y) = a(x,y) + b(x,y)\cos[\varphi(x,y) + 4\pi/3].$$

Here $a(x,y)$ and $b(x,y)$ represent the background illumination signal and the fringes' contrast function, respectively. The searched modulating phase is proportional to the profile of the 3D object under study plus a known term; therefore, it is trivial to solve for $p(x,y)$ once we find $\varphi(x,y)$ using Eq. (6) and (7).

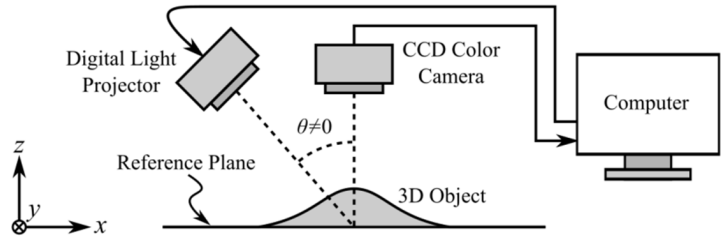

Fig. 2. Schematic of a single-camera single-projector setup for fringe projection profilometry.

Using RGB color encoding (see Fig. 3), we can project a single full-color image and generate simultaneously the three phase-shifted fringe patterns in Eq. (10). Without loss of generality we assume $f_0(x,y)$, $f_1(x,y)$, and $f_2(x,y)$ being projected in the red, green and blue channel, respectively. In absence of crosstalk distortion (which is feasible using dichromic filters and multiple sensors such as the ones inside of a 3-CCD camera), we can recover the searched fringe patterns from the recorded data in the RGB channels as [1, 9]:

$$\hat{I}_0(x,y) = I_R(x,y), \quad \hat{I}_1(x,y) = I_G(x,y), \quad \hat{I}_2(x,y) = I_B(x,y). \tag{11}$$

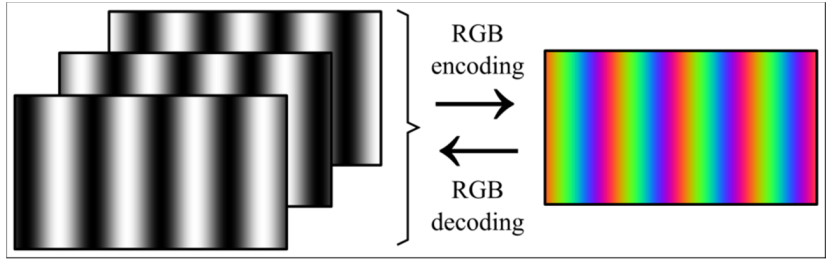

Fig. 3. RGB multichannel operation allows us to encode/decode up-to three phase-shifted open-carrier fringes into a single full-color image. Here we show low-frequency fringes for illustrative purposes; a much higher spatial frequency is used on the actual experiments.

It is a well-known fact that most single-sensor color cameras have overlapping sensibilities [9]. Therefore, when using this kind of cameras for fringe projection profilometry, RGB crosstalk distorts the recorded data and some compensation algorithm is required [2-6]. In practice, assuming the crosstalk phenomenon to be linear, we have:

$$\begin{bmatrix} I_R(x,y) \\ I_G(x,y) \\ I_B(x,y) \end{bmatrix} = \begin{bmatrix} A_{11} & A_{12} & A_{13} \\ A_{21} & A_{22} & A_{23} \\ A_{31} & A_{32} & A_{33} \end{bmatrix} \begin{bmatrix} I_0(x,y) \\ I_1(x,y) \\ I_2(x,y) \end{bmatrix}. \tag{12}$$

The real-valued matrix elements $A_{mn}$ represent the camera's sensibility on the $m$-th channel (for $m = \{R,G,B\}$) to a signal projected in the $n$-th channel ($n = \{R,G,B\}$).

Huang et al. [2], proposed a calibration method for matrix $\mathbf{A}$, which was later improved by Zhang et al. [3]. This method consists in generating only red, only green, and only blue phase-shifted open fringes and project them sequentially (instead of simultaneously) over a white reference plane. Due to crosstalk, the CCD camera records these single-color projected fringes on all three RGB channels. Analyzing separately each channel of the recorded data, we can easily calibrate all nine crosstalk coefficients $A_{mn}$ proportionally to the average of the fringes' contrast function [2,3]. Once we calibrate matrix $\mathbf{A}$, we can easily solve Eq. (12) for the crosstalk-compensated phase-shifted fringe-patterns as:

$$\begin{bmatrix} \hat{I}_0(x,y) \\ \hat{I}_1(x,y) \\ \hat{I}_2(x,y) \end{bmatrix} = \begin{bmatrix} B_{11} & B_{12} & B_{13} \\ B_{21} & B_{22} & B_{23} \\ B_{31} & B_{32} & B_{33} \end{bmatrix} \begin{bmatrix} I_R(x,y) \\ I_G(x,y) \\ I_B(x,y) \end{bmatrix}; \quad \text{where} \quad \mathbf{B} = \hat{\mathbf{A}}^{-1}. \tag{13}$$

We can introduce these crosstalk-compensated fringe patterns into Bruning's 3-step PSA:

$$A_0(x,y)\exp[i\hat{\varphi}(x,y)] = \begin{bmatrix} 1 & e^{-i2\pi/3} & e^{-i4\pi/3} \end{bmatrix} \begin{bmatrix} \hat{I}_0(x,y) \\ \hat{I}_1(x,y) \\ \hat{I}_2(x,y) \end{bmatrix} = \sum c_n \hat{I}_n(x,y). \tag{14}$$

However, from Eq. (13)-(14), it is easy to see that an equivalent but computationally more efficient approach to demodulate the searched analytic signal is given by:

$$A_0(x,y)\exp[i\hat{\varphi}(x,y)] = d_0 I_R(x,y) + d_1 I_G(x,y) + d_2 I_B(x,y);$$

$$\begin{bmatrix} d_0 & d_1 & d_2 \end{bmatrix} = \begin{bmatrix} 1 & e^{-i2\pi/3} & e^{-i4\pi/3} \end{bmatrix} \begin{bmatrix} B_{11} & B_{12} & B_{13} \\ B_{21} & B_{22} & B_{23} \\ B_{31} & B_{32} & B_{33} \end{bmatrix}. \quad (15)$$

The complex-valued coefficients $d_n$ need to be computed only once for fixed experimental conditions. The above equation requires 4 operations per pixel and it works directly over the raw data $\{I_R, I_G, I_B\}$. On the other hand, Eq. (14) required 16 operations per pixel and memory allocation for the intermediate results $\{\hat{I}_0, \hat{I}_1, \hat{I}_2\}$. Since RGB fringe-projection profilometry is meant for dynamic measurements [1-6], this represents a substantial reduction on the computational workload. As far as we know, Eq. (15) has never been reported before so this is our first contribution on this paper.

### 3. Influence of systematic errors on the phase-demodulation algorithm

Here we illustrate how systematic errors produce low-quality phase estimations when applying Bruning's 3-step PSA under non-ideal conditions, such as residual errors during the calibration of the crosstalk coefficients and distorting harmonics.

We start by modeling our estimation for the crosstalk matrix as $\hat{\mathbf{A}} = \mathbf{A} + \boldsymbol{\delta}$, where $\mathbf{A}$ describes the actual crosstalk values (see Eq. (12)), $\boldsymbol{\delta}$ is unknown, and for small errors we assume $|\delta_{mn}| \ll 1$. From Eq. (13)-(15), taking $\mathbf{BI}_{RGB} = \mathbf{B}(\mathbf{AI}_n) = \mathbf{B}(\hat{\mathbf{A}} - \boldsymbol{\delta})\mathbf{I}_n$, we have:

$$A_0(x,y)\exp(i\hat{\varphi}) = \sum c'_n I_n(x,y); \quad \begin{bmatrix} c'_0 \\ c'_1 \\ c'_2 \end{bmatrix}^T = \begin{bmatrix} 1 \\ e^{-i2\pi/3} \\ e^{-i4\pi/3} \end{bmatrix}^T \begin{bmatrix} 1+\varepsilon_{11} & \varepsilon_{12} & \varepsilon_{13} \\ \varepsilon_{21} & 1+\varepsilon_{22} & \varepsilon_{23} \\ \varepsilon_{31} & \varepsilon_{32} & 1+\varepsilon_{33} \end{bmatrix}, \quad (16)$$

where $\boldsymbol{\varepsilon} = -\mathbf{B}\boldsymbol{\delta}$; therefore $\varepsilon_{mn}$ are unknown, and $|\varepsilon_{mn}| < 1$ because $|\delta_{mn}| \ll 1$. Moreover, we know from Eq. (5) that the coefficients $c'_n$ define univocally not only the phase-demodulation formula but also the FTF: $H(\omega) = \sum c'_n \exp(-in\omega)$. In Fig. 4 we show several computer-simulated realizations of these FTFs, where $\varepsilon_{mn}$ are uniformly distributed random numbers within $(-0.15, 0.15)$. We must remark that these plots still correspond to Bruning's 3-step PSA but being applied to fringe patterns distorted by RGB crosstalk.

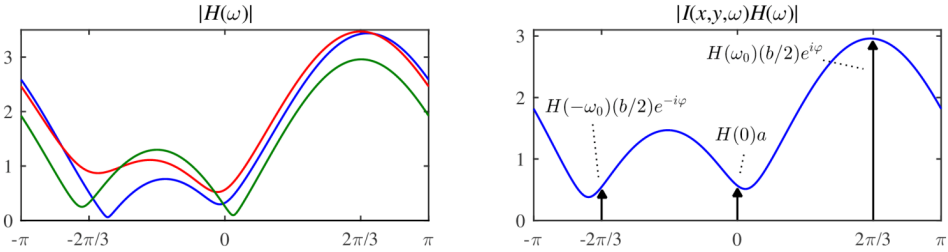

Fig. 4. Left side shows three FTFs given by $H(\omega) = \sum c'_n \exp(-in\omega)$ using uniformly distributed random values $\varepsilon_{mn} \in (-0.15, 0.15)$; each color represents a different realization. Right side shows the Fourier spectrum of the analytic signal estimated using one of these FTFs.

From Fig. 4 we see that $H(\omega) = \sum c'_n \exp(-in\omega)$ fails to fulfill the quadrature conditions, so the estimated analytic signal is given by:

$$A_0(x,y)e^{i\hat{\varphi}(x,y)} = H(0)a(x,y) + \frac{1}{2}b(x,y)\left\{H(2\pi/3)e^{i\varphi(x,y)} + H(-2\pi/3)e^{-i\varphi(x,y)}\right\}. \quad (17)$$

Carefully controlled experimental conditions and/or fringe pattern normalization allow us to eliminate the background signal term [13]. Thus, taking $a(x,y) \to 0$, we have:

$$A_0(x,y)e^{i\hat{\varphi}(x,y)} = \frac{1}{2}H(2\pi/3)b(x,y)e^{i\varphi(x,y)}\left\{1 + \frac{H(-2\pi/3)}{H(2\pi/3)}e^{i2\varphi(x,y)}\right\}. \quad (18)$$

For the trained eye, Eq. (18) predicts that the wrapped phase will be distorted by a double-frequency spurious signal, which is identical to the well-known distortion due to detuning error in phase-shifting interferometry [12]. We performed a numeric simulation to assess this pseudo-detuning distortion (see Fig. 6). First, we modulated 3 phase-shifted fringe patterns with a known phase. Next we estimated the analytic signal according to Eq. (16) using random values for the crosstalk calibration uncertainty, such as $|\varepsilon_{mn}| \leq \{0.05, 0.10, 0.15\}$. Finally we computed the wrapped difference between the actual phase and the estimated one, revealing the predicted double-frequency spurious signal. For these simulation, we used a low-frequency spatial carrier for ease of observation.

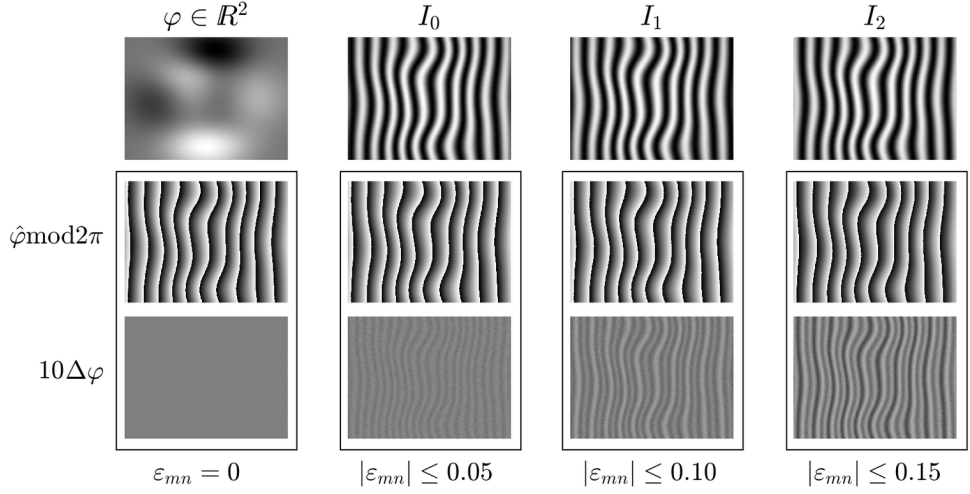

Fig. 6. The first row shows the phase map (given by MATLAB's peaks function) modulating three open-fringes phase-shifted interferograms. The second row shows the estimated phase as obtained from the 3-step PSA for different amounts of crosstalk-calibration error. The third row shows the wrapped difference $\Delta\varphi = (\varphi - \hat{\varphi}) \mod 2\pi$, multiplied by a factor 10 for ease of observation, revealing the predicted double-frequency ripple pattern. The intensity scale for the wrapped phases follows the standard linear mapping from $-\pi$ (black) to $\pi$ (white).

Up to this point, we have modeled perfectly sinusoidal fringe patterns for ease of exposition, but in practice these fringe patterns are distorted by harmonics:

$$I(x,y,t) = \sum_{n=-N}^{N} b_n(x,y) \exp\{in[\varphi(x,y) + (2\pi/3)t]\},$$

$$I(x,y,\omega) = \sum_{n=-N}^{N} b_n(x,y) \exp[in\varphi(x,y)]\delta(\omega - \omega_n); \quad \omega_n = \arg[\exp(i2n\pi/3)]. \quad (19)$$

Here $N$ represents the order of the highest harmonic with still-significant distorting energy, $b_n(x,y)$ is the contrast function of the $n$-th harmonic, and the wrapped location of the Dirac deltas is due the $2\pi$ − periodicity of the discrete-time Fourier transform [12]. When the fringe pattern is distorted by harmonics, we want to isolate $b_1 \exp(i\varphi)$ in order to find the best-quality estimation of the modulating phase. However, since $H(\omega) = \sum c'_n \exp(-in\omega)$ fails to fulfill the quadrature conditions $H(-2\pi/3) = H(0) = 0$, we have:

$$\begin{aligned}
I(\omega)H(\omega) = & H(2\pi/3)\{b_1 e^{i\varphi} + b_2 e^{-i2\varphi} + b_4 e^{i4\varphi} + b_5 e^{-i5\varphi} + ...\}\delta(\omega - 2\pi/3) \\
& + H(0)\{b_0 + b_3 e^{i3\varphi} + b_3 e^{-i3\varphi} + b_6 e^{i6\varphi} + b_6 e^{-i6\varphi} + ...\}\delta(\omega) \\
& + H(-2\pi/3)\{b_1 e^{-i\varphi} + b_2 e^{i2\varphi} + b_4 e^{-i4\varphi} + b_5 e^{i5\varphi} + ...\}\delta(\omega + 2\pi/3).
\end{aligned} \quad (20)$$

It is easy to see from the above equation that back on the temporal domain we will have an analytic signal distorted by pseudo-detuning (because $b_1 \exp(-i\varphi)$ remains unfiltered), and corrupted by the distorting harmonics with the highest energy (both components of the second one, both components of the third one, and so on). Therefore, it is clear that we require a more robust approach to cope with non-sinusoidal fringes and severe RGB crosstalk.

**4. Robust quadrature filtering for crosstalk-distorted fringe patterns.**

As we know from the FTF formalism (see Appendix A in [12]), the PSAs with detuning error robustness require four or more phase-shifted interferograms. Nevertheless, squeezing interferometry allows us to achieve detuning robust quadrature filtering and decrease the harmonics distortion using just three phase-shifted samples [10].

The first step in squeezing interferometry is to rearrange the phase-shifted fringe patterns ($\{\hat{I}_0, \hat{I}_1, \hat{I}_2\}$ in our case), in order to produce a spatial-carrier fringe pattern with very high spatial-carrier frequency as follows:

$$\left.\begin{aligned}
I_S(3x-2, y) &= \hat{I}_0(x,y) \\
I_S(3x-1, y) &= \hat{I}_1(x,y) \\
I_S(3x-0, y) &= \hat{I}_2(x,y)
\end{aligned}\right\} \quad (0,0) \leq (x,y) \leq (X,Y), \quad (21)$$

where $(X,Y)$ represent the horizontal and vertical sizes of $\hat{I}_n(x,y)$. Rewriting the resulting spatial-carrier fringe pattern in terms of $x' = 3x$, we have:

$$I_S(x', y) = \sum_{n=-N}^{N} b_n(x', y) \exp\{in[\varphi(x',y) + 2\pi x'/3]\}; \quad (0,0) \leq (x',y) \leq (3X,Y). \quad (22)$$

And computing its discrete Fourier transform results: $\mathcal{F}\{I_S(x',y)\} = I_S(u,v)$,

$$I_S(u,v) = \sum_{n=-N}^{N} \mathcal{F}\{b_n(x',y)\exp[in\varphi(x',y)]\} * \delta(u - u_n, v); \quad u_n = \arg[\exp(i2n\pi/3)]. \quad (23)$$

The asterisk represents the convolution operation, and we are applying the translation property of Dirac delta: $f(u,v) * \delta(u - \Delta, v) = f(u - \Delta, v)$. Factorizing the spectral lobes by their spectral displacement, we can rewrite the above equation as:

$$I_S(u,v) = \mathcal{F}\left\{b_1 e^{i\varphi} + b_2 e^{-i2\varphi} + b_4 e^{i4\varphi} + b_5 e^{-i5\varphi} + b_7 e^{i7\varphi} + ...\right\} * \delta(u - 2\pi/3, v)$$
$$+ \mathcal{F}\left\{b_0 + b_3 e^{i3\varphi} + b_3 e^{-i3\varphi} + b_5 e^{i6\varphi} + b_5 e^{-i6\varphi} + ...\right\} * \delta(u,v) \quad (24)$$
$$+ \mathcal{F}\left\{b_1 e^{-i\varphi} + b_2 e^{i2\varphi} + b_4 e^{-i4\varphi} + b_5 e^{i5\varphi} + b_7 e^{-i7\varphi} + ...\right\} * \delta(u + 2\pi/3, v).$$

Comparing Eq. (20) and (24), it is clear that the temporal and spatial carriers produce similar spectral distributions, but there are significant differences. First, the energy of $\mathcal{F}\{b_n e^{\pm in\varphi}\}$ spreads over a spectral region $n$ times as large as the bandwidth of $\mathcal{F}\{b_1 e^{i\varphi}\}$; thus, the $n$-th harmonic has much lower spectral density than the searched analytic signal. Also, processing in the spatial Fourier domain is immune to detuning error because we can filter-out all signals outside the area of interest, instead of just the components exactly at $(-2\pi/3, 0)$ and $(0,0)$. As illustrated in Fig. 5, this means we can achieve robust quadrature filtering with the following synchronous demodulation formula [10,11]:

$$A_0(x', y) \exp[i\hat{\varphi}(x', y)] = \exp(-i2\pi x'/3) \times QF\{I_S(x', y)\}. \quad (25)$$

Here $QF\{\cdot\}$ represents a narrow quadrature filter in the Fourier domain (usually a binary mask) which preserves only the first harmonic of the analytic signal. Once we estimate $A_0(x', y) \exp[i\hat{\varphi}(x', y)]$, we squeeze it back to the original size of $\hat{I}_n(x,y)$, and finally we compute its argument and unwrap the modulating phase, as usual [10,11].

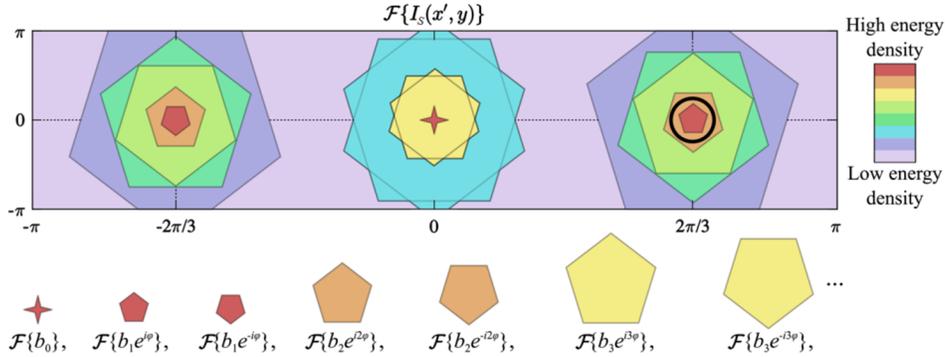

Fig. 5. Schematic illustration of Fourier demodulation for the spatial-carrier fringe pattern $I_S(u,v)$ in Eq. (24). The black circle represents a binary mask band-passing the searched analytic signal (red pentagon) and rejecting all unwanted distorting harmonic components.

## 5. Experimental results

In our experiment, we worked with a static test subject for repeatability: digitizing a human-face model made of expanded polystyrene (see Fig. 7). This allows us to take a crosstalk-free measurement using *sequential* grayscale-based fringe projection as null test. We used this null test to assess the quality of the recovered phases from single-shot RGB fringe projection using Bruning's 3-step PSA and squeezing interferometry. The open fringes were projected using a $1024 \times 768$ digital light projector, with defocusing and pre-encoding gamma [14,15]. The fringe patterns were captured using a single-sensor digital color camera that performs RGB color separation using a Bayer filter mosaic with a $1280 \times 768$ resolution and 8-bit color depth per channel. We calibrated the crosstalk matrix using 6-step least-squares phase-shifting measurements of the reference plane for increased reliability (to generalize from [2,3] is straightforward). Our numerical values for the crosstalk calibration matrix $\hat{\mathbf{A}}$ (see Eq. (12)- (13)) and for the corresponding phase demodulation coefficients $d_n$, required to work directly with raw data (see Eq. (15)), are given by:

$$\hat{\mathbf{A}} = \begin{pmatrix} 0.4334 & 0.4041 & 0.0749 \\ 0.0791 & 0.9092 & 0.3316 \\ 0.0007 & 0.3679 & 0.9536 \end{pmatrix}, \begin{pmatrix} d_0 \\ d_1 \\ d_2 \end{pmatrix} = \begin{pmatrix} 2.2252 - i0.3585 \\ 0.4581 + i1.9615 \\ -0.8586 + i0.2542 \end{pmatrix}. \quad (26)$$

According to this numerical values, we faced severe RGB-crosstalk distortion in comparison to other calibration matrices reported in the literature [2,3,6].

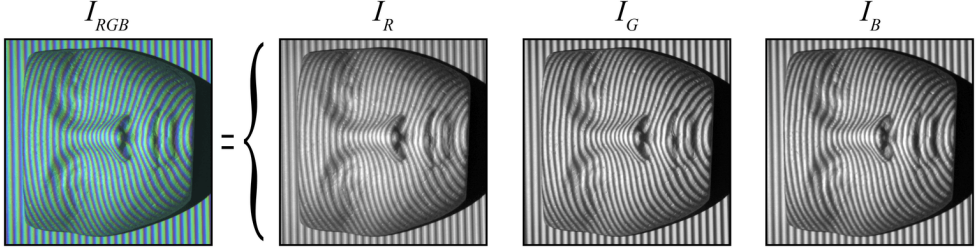

Fig. 7. Photograph of our test subject being illuminated with RGB phase-shifted open fringes.

Figure 8 shows the first step of the squeezing interferometry method [10], where we rearrange the information from the phase-shifted fringe patterns $\{\hat{I}_0(x,y), \hat{I}_1(x,y), \hat{I}_2(x,y)\}$ in order to generate the very high-frequency spatial-carrier fringe pattern $I_S(x',y)$. The beating distortion observed in $I_S(x',y)$ is a presentation artifact (nonexistent in the actual data) as consequence of the down sampling from the original size ($3840 \times 768$ pixels).

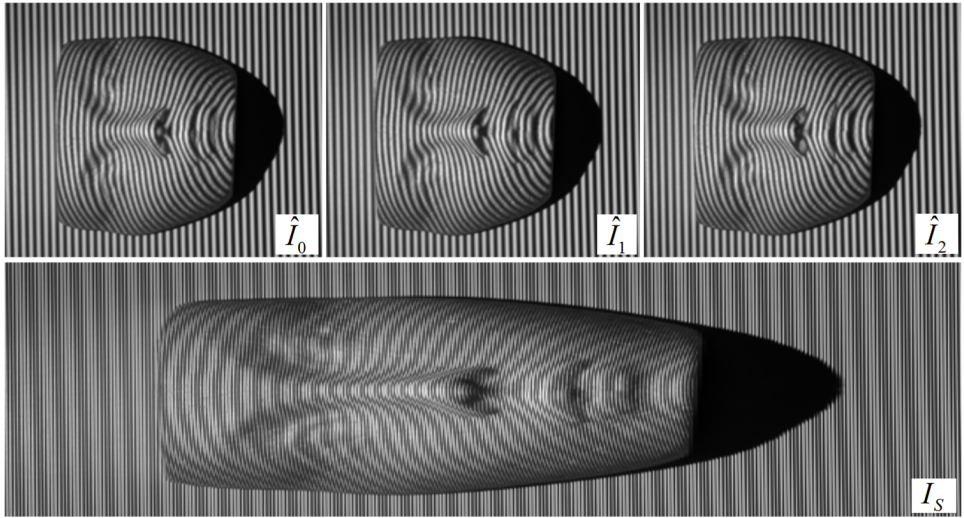

Fig. 8. First stage of the squeezing interferometry method as described in Eq. (21).

The next step is to compute $\mathcal{F}\{I_S(x',y)\}$, from where we isolate the searched analytic signal by means of a narrow one-sided bandpass filter. Back on the spatial domain, we squeezed the analytic signal back to the original size of $I_n(x,y)$, and then we estimated the wrapped phase from its argument. For ease of visualization, at this point we removed the low-frequency modulation $u_0 x \tan(\theta)$, previously calibrated from the reference plane. See Fig. 9.

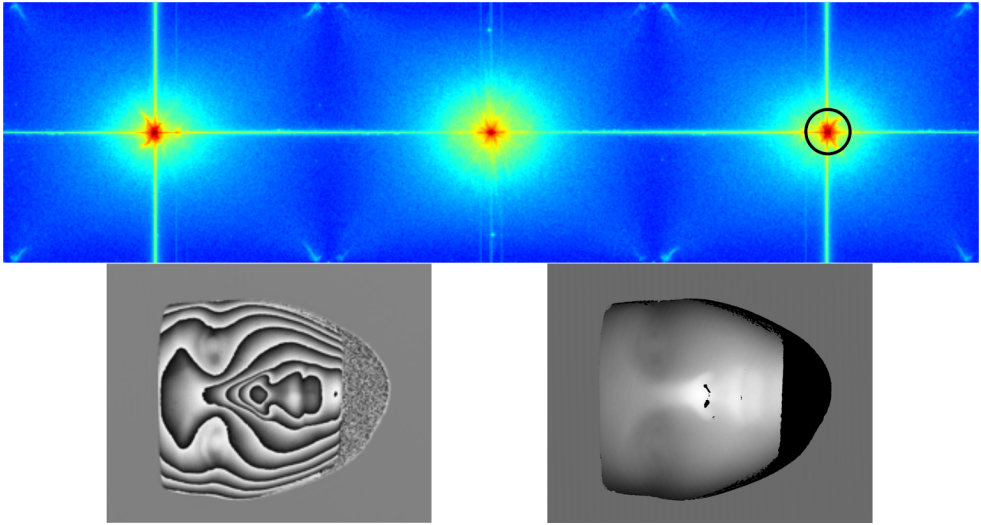

Fig. 9. Last steps of the squeezing interferometry method: quadrature filtering of $\mathcal{F}\{I_S(x',y)\}$, wrapped phase estimated as the argument of the analytic signal inside the black circle, and the unwrapped phase proportional to the profile of our test subject.

Figure 10 shows the recovered profiles from the RGB raw data, the crosstalk compensated fringe patterns, squeezing interferometry, and our null-test for illustrative purposes. These plots use the amplitude of the analytic signal as texture in order to visualize fine details.

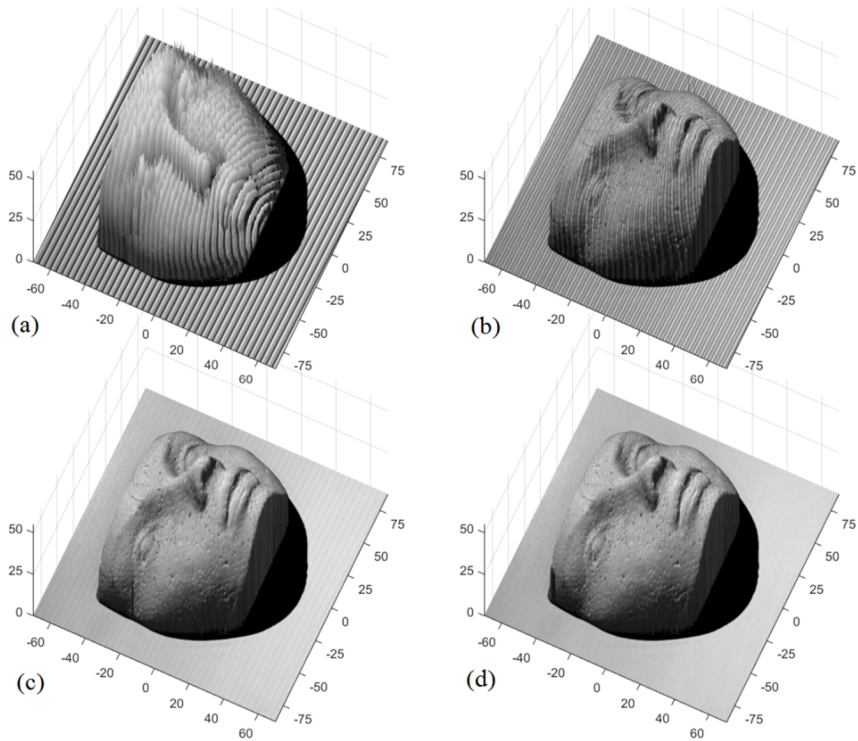

Fig. 10. Tridimensional render of the digitized object with RGB fringe projection profilometry. These panels are described in the main text. Units of all axes are in millimeters.

Figure 10(a) shows the 3D profile recovered from the raw camera data, where crosstalk-distortion was so severe that the phase unwrapper failed to recover even the overall shape of the test subject. Figure 10(b) shows the result of processing the same raw data using our proposed algorithm (Eq. (15)); this result shows a drastic improvement respect to Fig. 10(a) but still exhibits the ripple pattern due to pseudo-detuning error and distorting harmonics. Figure 10(c) shows our best quality result, which we obtained with the combination of crosstalk-compensation and squeezing interferometry. Finally, Fig. 10(d) shows the result of our crosstalk-free null test for illustrative purposes; this result was obtained by means of *sequential* gray-scale based fringe projection.

Figure 11 shows horizontal slices of the unwrapped phase for qualitative comparison of the results estimated with Bruning's 3-step PSA and with squeezing interferometry. This figure also demonstrates that the ripple pattern due to pseudo-detuning and distorting harmonics is observed on the modulated phase.

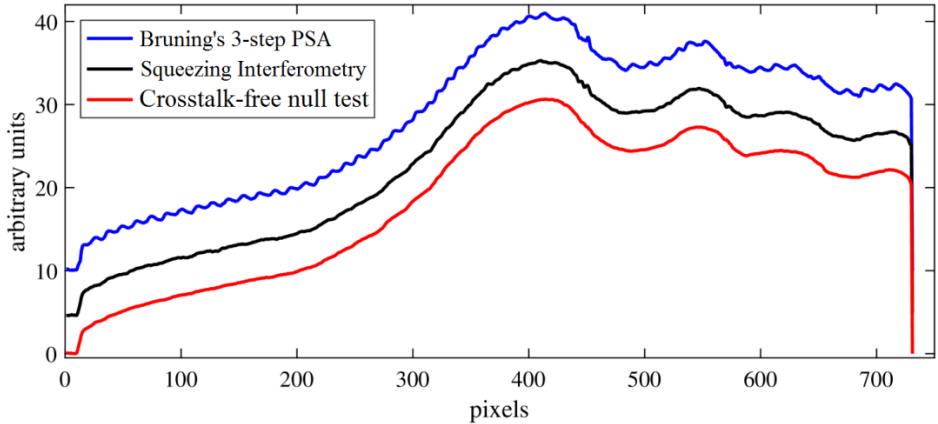

Fig. 11. Horizontal slices of the unwrapped phase estimated with Burning's 3-step PSA, the squeezing interferometry method, and the crosstalk-free null test. A global piston was included between each plot for ease of visualization.

## 6. Conclusions

We analyzed the most commonly used phase-demodulation approaches for RGB fringe-projection profilometry by means of the FTF formalism. We proposed a computationally more efficient algorithm applicable for mild to moderate RGB-crosstalk conditions, which combines crosstalk compensation and phase demodulation, and that works directly with the raw data (Eq. (15)). We show that relying on Bruning's 3-step PSA for severe crosstalk conditions (when modeling this phenomenon using linear combinations no longer applies) translates into a pseudo-detuning distortion and high-energy distorting harmonics. The pseudo-detuning error was somehow unexpected because the phase-step is a controlled parameter in fringe projection techniques (see Eq. (9)). We also analyzed the squeezing interferometry phase-demodulation method to show that it allows us to recover high-quality results even under severe crosstalk distortion. Finally, we supported our claims with numerical simulations and experimental results.

**Acknowledgments**

The authors want to thank the financial support from the Mexican National Council for Science and Technology (CONACYT), grant CB2010-157044. Also the authors acknowledge Cornell University for supporting the e-print repository arXiv.org and the Optical Society of America for allowing the contributors to post their manuscript at arXiv.